\documentclass[12pt]{iopart}
\pdfoutput=1
\usepackage{url}
\usepackage{booktabs}
\usepackage{subfigure}
\usepackage[load-configurations=abbreviations,per-mode=symbol]{siunitx}
\usepackage{graphicx} 
\newcommand{\eg}{e.g., }
\newcommand{\ie}{i.e., }

\begin{document}

\title[Reproducibility of spatiotemporal traffic dynamics]
	{On the reproducibility of spatiotemporal traffic dynamics with microscopic traffic models} 

\author{Florian Knorr and Michael Schreckenberg}

\address{Fakult\"at f\"ur Physik, Universit\"at Duisburg-Essen, 47048 Duisburg, Germany}
\eads{\mailto{knorr@ptt.uni-due.de}}

\begin{abstract}
Traffic flow is a very prominent example of a driven non-equilibrium system.
A characteristic phenomenon of traffic dynamics is the spontaneous and abrupt 
drop of the average velocity on a stretch of road leading to congestion. 
Such a traffic breakdown corresponds to a boundary-induced phase 
transition from free flow to congested traffic.
In this paper, we study the ability of selected microscopic traffic models to 
reproduce a traffic breakdown, and we investigate its spatiotemporal dynamics. 
For our analysis, we use empirical traffic data from stationary loop detectors on 
a German Autobahn showing a spontaneous breakdown.
We then present several methods to assess the results and compare the models 
with each other. 
In addition, we will also discuss some important modeling aspects and their impact 
on the resulting spatiotemporal pattern. 
The investigation of different downstream boundary conditions, for example, shows that the physical 
origin of the traffic breakdown may be artificially induced by the setup of the boundaries. 
\end{abstract}

\pacs{
05.10.-a, 
64.60.De, 
89.75.-k, 
89.40.Bb  
}
\maketitle

\section{Introduction}
\label{sec:introduction}
During the past 50 years, numerous models to describe vehicular traffic have been 
proposed (see, \eg the review articles~\cite{ChowdhurySantenSchadSchneider2000,Helbing2001,Nagatani2002,NagelWagnerWoesner2003}). 
Based on the approach to the complex many-body system traffic, 
these models can be further classified as macroscopic or microscopic.
Macroscopic models treat traffic flow analogously to the flow of a compressible fluid 
and, hence, are restricted to study the collective dynamics instead of the individual vehicles' motion. 
Microscopic models, on the other hand, explicitly model vehicle-vehicle interactions and 
keep track of every single vehicle. 
Depending on the area of application, either approach 
may be favored---possible criteria for the evaluation are the desired level of detail, the computational tractability, or 
their ability to reproduce particular empirical features of traffic flow.

The pursuit of finding better and better models has led to a multitude of macroscopic and microscopic traffic models.
In the field of microscopic models alone, one currently counts more than one hundred such models 
\cite{BrockfeldKuehneSkabardonisWagner2003}, and their number is still increasing. 
These models, however, may be subdivided again.
Traffic cellular automata \cite{MaerivoetDeMoor2005}, in which both space 
and time are discrete variables, represent a prominent subclass of microscopic traffic models. 
Together with their rule-based dynamics, they are closely related to the particle hopping models known from 
non-equilibrium physics (\eg TASEP or ZRP \cite{DerridaEvans1997, SchadschneiderChowdhuryNishinari2010}). 
In contrast, car-following models are continuous in space and time. In this subclass, a vehicle's motion 
is governed by a differential equation---usually involving the position, the velocity, and the acceleration 
of the preceding vehicle (for a comprehensive overview of the various modeling approaches, see 
\cite{SchadschneiderChowdhuryNishinari2010}). 

The question of what characterizes a `good' model is not clear though 
as several levels of detail can be considered: inter-vehicle 
dynamics (how a vehicle interacts with its immediate predecessor), traffic dynamics (whether the model 
exhibits the known traffic patterns), or traffic statistics (whether the model is able to reproduce 
empirical lane usage and headway distributions). This distinction is necessary as a model performing
well in one of these fields is not guaranteed to behave equally well in the others.

Despite the large number of models, there are relatively few studies which compare these models 
to empirical traffic data. Especially in the field of traffic dynamics, the authors found a lack of 
comparative studies, which was the starting point of this work.

The fundamental observables of traffic dynamics are traffic flow $J$, average velocity $v$ and 
vehicle density $\rho$. 
The functional relation between vehicle density and traffic flow is often referred to as the fundamental diagram.
However, the fundamental diagram, combining temporally 
aggregated and averaged data, hides the vehicle dynamics and allows only for a coarse-grained analysis.

It is a very interesting question whether traffic models are able to predict the transition from 
one traffic phase to another. Investigating this question is challenging as such a phase transition occurs
spontaneously and is not restricted to a certain location on the road. Moreover, it is not clear how 
to assess the performance of a model compared with empirical traffic data.

In this paper, we want to discuss the ability of three selected traffic models to reproduce 
the spatiotemporal dynamics of traffic flow. To assess the quality of the results, we will 
present methods that allow both a qualitative and a quantitative analysis. 
In this context, we will also discuss some modeling aspects and their influence on the 
observed traffic pattern. We modeled a section of a German Autobahn and 
used empirical detector data that show a phase transition in the morning peak hour.
As we explicitly considered multilane dynamics in heterogeneous traffic, we chose models 
with asymmetric lane changing rules to mimic the lane changing behavior that is found 
not only on a German Autobahn but also in most other European countries.

The rest of this paper is structured as follows. In section \ref{sec:related_work} we give an overview of related 
work investigating the properties of microscopic traffic models with respect to real world measurements.
In section \ref{sec:data_and_testing} we introduce several methods to assess the quality of the simulation 
results with respect to the empirical data. Section \ref{setup} gives an overview of the microscopic traffic 
models used in this work and describes the simulation setup. A detailed description of the setup---especially 
the setup of boundaries and ramps---is necessary as it can have a strong influence on the results and their reproducibility.
Before summarizing our findings in section \ref{sec:conclusion}, we present our results in section \ref{sec:results}.

\section{Related work}
\label{sec:related_work}
Vehicular traffic is a system showing very complex behavior (\eg metastability, shock-wave formation 
and dynamic phase transitions~\cite{ChowdhurySantenSchadSchneider2000,Helbing2001,NagelWagnerWoesner2003}). 
Above a critical density, local inhomogeneities can trigger a collective phenomenon: a traffic breakdown. 
The initial position of the breakdown is usually located at a bottleneck (\eg an on-ramp or an off-ramp) 
from where the congested traffic pattern propagates upstream. 

A similar behavior is known from one-dimensional driven particle systems with open boundaries. 
The bulk dynamics of such systems is governed by the rates at which particles enter or leave the system 
at the boundaries~\cite{KolomeiskySchuetzKolomeiskyStraley1998,PopkovSchuetz1999,AntalSchuetz2000,
HagerKrugPopkovSchuetz2001}. 
The resulting phase diagram reveals distinct phases separated by first and second order phase 
transitions, respectively. Depending on the inflow and outflow rates of the system, a local perturbation 
may move either along the flow of particles or in the opposite direction. 
When compared with vehicular traffic, the latter case may be interpreted as a traffic jam propagating upstream. 
Similarly, the shock, which marks a discontinuity in the density profile, can be seen as the jam's upstream front.
Hence, a traffic breakdown is a spatiotemporal phenomenon, whose 
observation requires a relatively broad spatial and temporal horizon.

Currently, stationary loop detectors are still the most common source of traffic data. In their 
simplest form they count the number of passing vehicles and measure their velocity aggregated 
over intervals of one minute. These values
already allow an empirical fundamental diagram of traffic flow to be drawn. 
A more detailed picture on inter-vehicle dynamics can be obtained if even single vehicle data are available. 
Knospe~\etal~\cite{KnospeSantenSchadschneiderSchreckenberg2004}, for instance, 
used data from loop detectors which also measured time-headways (\ie the time passing between two 
vehicles crossing a detector). They analyzed the distribution of time-headways 
depending on the density and studied the functional relation between speed and
distance to the preceding car, known as optimal velocity function. In addition to that, they compared these 
data to seven traffic cellular automata (CA) models. 
Knospe \etal found significant differences between the examined models. 
In particular, the earlier and simpler models 
were not able to satisfyingly reproduce the empirical results. More advanced models like the comfortable driving 
model (CDM), which is one of the models to be studied in this paper, showed good agreement with the empirical data. 

As detector data represent locally aggregated information, they allow no explicit statement 
on the spatial extent of traffic states. 
%

For this reason the data of a single detector do not suffice to study the spatiotemporal dynamics.
Analyzing the time series of a sequence of neighboring detectors removes this restriction. Therefore, 
we have chosen a highway section with a sufficient number of detectors for our study. To examine spatiotemporal 
traffic dynamics, we have selected two models that gave good results in previous studies. As a reference, we have 
also included the Nagel-Schreckenberg model (NSM)~\cite{NagelSchreckenberg1992}, which is a rather simplistic 
traffic cellular automaton~\cite{MaerivoetDeMoor2005}. 

The selected models come with rules for asymmetric lane changing as required on the selected highway. 
Lane changes certainly are one potential source of local perturbation of traffic flow and have to be considered 
for a realistic reproduction of the scenario.

The only comparisons between empirically observed and simulated traffic 
dynamics that the authors are aware of were carried out by Treiber~\etal~\cite{TreiberHenneckeHelbing2000}, Popkov~\etal~\cite{PopkovSantenSchadschneiderSchuetz2001} and Kerner~\etal~\cite{KernerKlenovHiller2007}. 
The first two articles, however, focused
on single lane dynamics and did not provide quantitative results. Lane changes have 
an important influence on traffic dynamics, as Kerner and Klenov~\cite{KernerKlenov2009} found by analyzing
vehicle trajectories: lane changing between neighboring lanes is responsible for the emergence 
(and dissolution) of congested traffic states.
The article by Kerner~\etal offers a very detailed discussion of traffic dynamics and a qualitative comparison of 
empirical data with two models based on Kerner's three phase traffic theory. The authors, however, do not 
provide a quantitative analysis nor do they discuss the influence of the various parameters on their 
results. 

\section{Empirical data and model testing}
\label{sec:data_and_testing}
\begin{figure}
\centering
\includegraphics[angle=0,width=.8\textwidth]{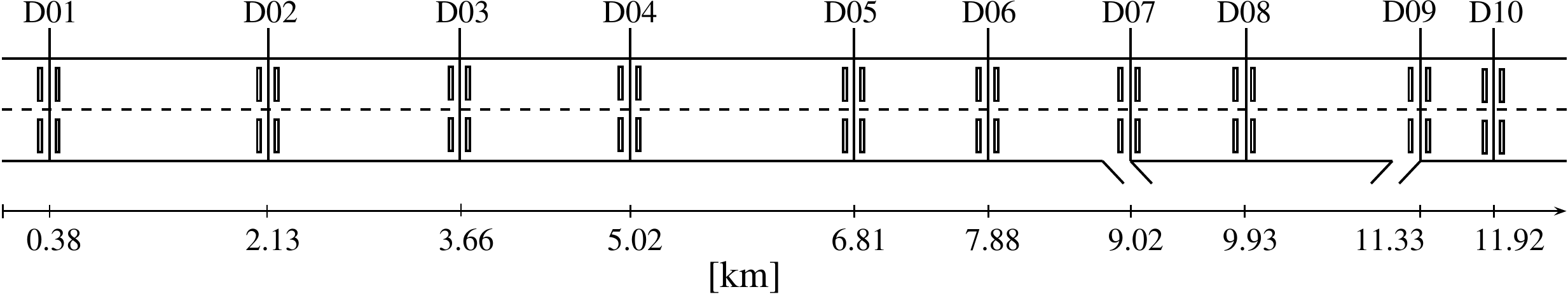}
\caption{Schematic sketch of the highway section considered. The detectors are labeled as D01,$\ldots$,D10.}
\label{fig:highway}
\end{figure}

To investigate the spatiotemporal traffic dynamics, we used data from ten detector cross-sections 
on the German autobahn A044 between the cities of Unna and Werl. A schematic sketch of the two-lane highway 
section is depicted in figure~\ref{fig:highway}. The section is well suited for our analysis as it contains 
a large number of detectors and single off- and on-ramps at its downstream end, which serve as bottlenecks. 
Both ramps can potentially trigger a traffic breakdown. The upstream cross-sections allow measurements 
of the traffic patterns generated at the bottlenecks without perturbations by additional bottlenecks. 

The detectors on this section distinguish two vehicle classes, namely ``trucks'' and ``cars''. 
(Yet there is no strict rule that this classification is based on.) 
For each vehicle class the detectors measure the number and 
the average velocity of all vehicles that pass the corresponding cross-section per one minute interval. 
The vehicle density $\rho$, of which a direct measurement is difficult, can 
be estimated from the average velocity $v$ and average traffic flow $J$ via the fundamental relation
\begin{equation}
J=v\rho.
\end{equation}
(Note that this estimate is only a good approximation in dilute traffic, whereas in dense traffic this 
approach tends to overestimate the density's actual value~\cite{KernerKlenovHiller2007}.)

\begin{figure}
\centering
\includegraphics[width=\textwidth]{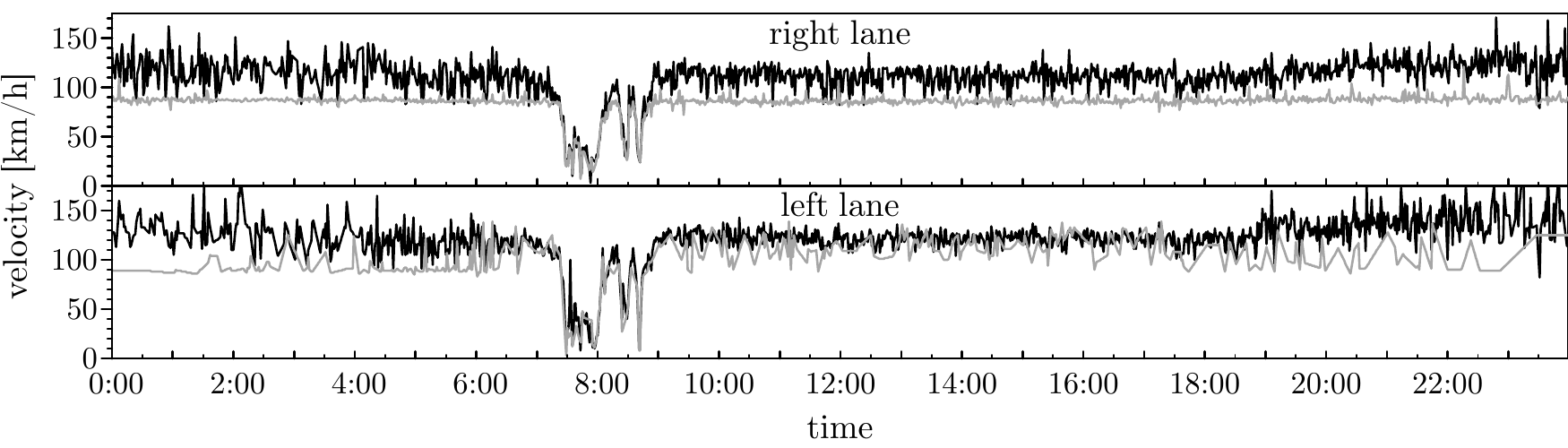}
\caption{The velocity time series of detector D05 separated by lane and vehicle type.
The time series of cars is depicted with a solid black line. 
The corresponding time series of trucks is depicted with a solid gray line. 
They are exemplary for all detectors upstream of
the bottleneck. The breakdown occurs at approximately 7:15~a.m., it affects both lanes, and it lasts 
approximately \SI{90}{\minute}. Microscopic traffic models should be able to reproduce this breakdown. 
(Missing data points, usually resulting from the lack of trucks in dilute traffic in the left lane, were
approximated by a linear interpolation for better readability.)}
\label{exemplary-time-series}
\end{figure}

For our analysis we used the data as detected on 4 November 2010. 
As one can see from figure~\ref{exemplary-time-series}, there was a spontaneous traffic breakdown during the morning 
peak hour on this day. 

By aggregating the same information as the real 
detectors, the simulations yield an equivalent set of data. 
To compare the resulting data, we propose two very intuitive approaches. Let 
$V_\mathrm{e/s}=\left\{v_\mathrm{e/s}(t_1),v_\mathrm{e/s}(t_2),...,v_\mathrm{e/s}(t_n)\right\}$
signify the empirical (subscript e) and the simulated (subscript s) time series of a detector. 
By regarding the two sequences of length $n$ as $n$-dimensional vectors Brockfeld~\etal~\cite{BrockfeldWagner2005} 
suggested using the 1-norm $L_1$
\begin{equation}
\label{eqn:1norm}
L_1 = \sum_{i=1}^n |v_\mathrm{e}(t_i)-v_\mathrm{s}(t_i)|
\end{equation}
as a direct error measure. 
For the similarity analysis of two time series, however, the shape of the series is, in general, more important than their 
absolute values. Hence, random fluctuations in free traffic flow should be filtered and should not enter the comparison. 
Moreover, as the traffic models use different upper boundaries for vehicle velocities 
(see table~\ref{table-model-compare}), the resulting time series might have different amplitudes. 
Therefore, and according to 
\cite{GoldinKanellakis1995}, we suggested a normalization of the series before calculating the norm. 
The normalization of a time series $X=\left\{x_1,x_2,...,x_n\right\}$ with mean $\mu(X)$ 
and standard deviation $\sigma(X)$ is obtained 
by the affine transformation
\begin{equation}
x_i \rightarrow \frac{x_i - \mu(X)}{\sigma(X)}.
\label{eqn:affine-transformation}
\end{equation}
One should mention that the 1-norm is just one way of determining the distance between two time series. In general, any 
symmetric and positive-definite metric for which the triangle inequality holds could be applied.

Another promising method to assess the models' quality is the study of the time series' residuals. The residuals $R(t_i)$ denote 
the difference between the observed and the simulated value at each time step $t_i$
\begin{equation}
R(t_i) = v_\mathrm{e}(t_i) - v_\mathrm{s}(t_i).
\label{eqn:residual}
\end{equation}
A classical approach to analyze time series is the decomposition of the series into three components \cite{Madsen2007}:
a trend component, a seasonal component, and a noise component.
If the model can reproduce the empirical data, the calculation of the residuals cancels the trend and seasonal 
components and leaves only the noise term. Under the assumption of white noise, the residuals are expected to behave randomly and to be 
uncorrelated. Similarly, it is also possible to determine the correlation between the two series. In this case, the degree to which 
the simulated time series reproduces the empirical data can be determined by the cross correlation of the two series.

\section{Simulation setup}
\label{setup}
\subsection{Models tested}
For our analysis we restricted ourselves to the following three microscopic models 
for which asymmetric passing rules already exist: the Nagel-Schreckenberg model (NSM)~\cite{NagelSchreckenberg1992},
the comfortable driving model (CDM)~\cite{KnospeSantenSchadschneiderSchreckenberg2000,KnospeSantenSchadschneiderSchreckenberg2002}, 
and the intelligent driver model (IDM)~\cite{TreiberHenneckeHelbing2000}. 

The NSM and CDM are both traffic cellular automata (CA), where space and time are discrete, as explained earlier. 
It should be noted, though, that the newer CDM uses a finer discretization. 
In addition, the CDM was extended by some anticipatory components, which enable a vehicle 
to react more carefully to the preceding vehicle. The NSM, on the other hand, ignores any preceding 
vehicle, unless a collision is imminent. 
The IDM, in contrast, is a car-following model in continuous space and time. 
By a suited adaption to the preceding vehicle, the underlying differential equation leads 
to a very smooth driving behavior with realistic values of acceleration and deceleration. 
As the numerical evaluation of the IDM also requires a temporal discretization, the value of the temporal 
discretization in table \ref{table-model-compare} refers to the discretization  that we used in 
our simulations. The NSM's values were taken from \cite{NagelSchreckenberg1992,KnospeSantenSchadschneiderSchreckenberg1999}.
We did, however, double the length of slow vehicles to better mimic the physical properties of trucks.
Analogously, we doubled the length of slow vehicles in the CDM. In contrast to Knospe \etal 
\cite{KnospeSantenSchadschneiderSchreckenberg2000,KnospeSantenSchadschneiderSchreckenberg2002}, 
we augmented the maximum velocity of both cars and trucks to obtain more realistic values with respect to a German Autobahn, 
but the absolute difference in maximum velocities was preserved.
\begin{table}
	\caption{Comparison of the mobility models and the properties of vehicles in the respective model. 
	In the case of asymmetric lane usage the IDM requires an additional parameter to model longitudinal motion. 
	(The number in brackets refers to the symmetric case.) IDM's parameters are taken from \cite{KestingTreiberHelbing2007}.}
	\label{table-model-compare}
	\centering
	\begin{tabular}{lrrrr}
		\toprule  										& unit& NSM & CDM 	& IDM   \\
			\midrule
			spatial discretization &[\si{\meter}]			& $7.5$ & $1.5$ & $0.0$ \\ 
			temporal discretization &[\si{\second}] 		& $1.0$	& $1.0$	& $0.25$\\ 
			max. velocity car &[\si{\meter\per\second}] 	& $37.5$& $33.0$& $34.0\pm 20\%$\\ 
			max. velocity truck &[\si{\meter\per\second}] 	& $22.5$& $25.5$& $23.0\pm 20\%$\\ 
			max. acceleration &[\si{\meter\per\square\second}]& $7.5$& $1.5$ & $1.5$ \\
			car length &[\si{\meter}]						&  $7.5$& $7.5$ & $4.0$ \\
			truck length &[\si{\meter}]						& $15.0$& $15.0$& $12.0$\\
			additional parameters&							& $1$	& $4$	& $5(4)$\\
		\bottomrule
	\end{tabular} 
\end{table}

\begin{figure}
	\centering
	\includegraphics[width=.5\textwidth]{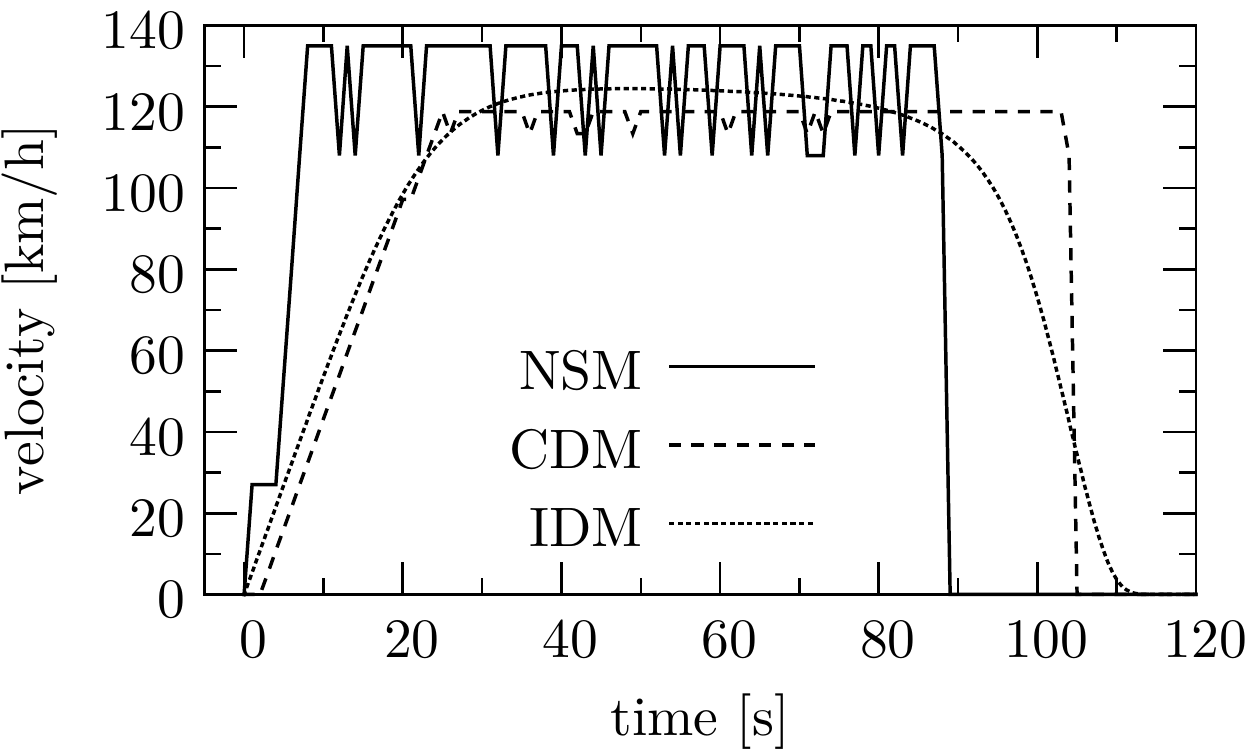}
	\caption{Velocity profiles for a vehicle starting from rest with a standing vehicle \SI{3}{\km} ahead 
	for the NSM, CDM, and IDM.}
  \label{fig:velocity-profile}
\end{figure}

The single vehicle dynamics reveals the differences between the models. 
For each model, figure~\ref{fig:velocity-profile} shows the velocity profile of a vehicle starting at rest and approaching
a parked vehicle \SI{3}{\kilo\meter} ahead. The discretization of the CA 
manifests itself in discontinuities of the corresponding velocity profile. The delayed acceleration
in the CDM is due to a so-called slow-to-start rule. 
Due to this rule, a vehicle accelerates only with a given probability, when starting from rest. 
When approaching the parked vehicle, the IDM's vehicle initiates a smooth braking process with realistic deceleration values. 
In the CA models, the vehicle comes to rest within one or two time steps 
(corresponding to \SIrange[range-phrase = --]{1}{2}{\second}) after driving at maximum speed. 
It has to be noted, however, that the resulting high deceleration rates can be observed in all models on roads with  
multiple lanes and on-ramps.

The asymmetric lane changing rules for the NSM are described in 
\cite{RickertNagelSchreckenbergLatour1996,KnospeSantenSchadschneiderSchreckenberg1999}. 
(Several other asymmetric rule sets have been proposed. For an overview see the review article~\cite{ChowdhurySantenSchadSchneider2000}.)
The CDM's model description~\cite{KnospeSantenSchadschneiderSchreckenberg2000,KnospeSantenSchadschneiderSchreckenberg2002} 
also provides the corresponding set of lane changing rules. A lane changing model for the 
IDM, called MOBIL, is given in~\cite{TreiberKesting2007,KestingTreiberHelbing2007}.
Referring to the cited articles, we will skip a detailed review of 
the models in favor of a precise description of the simulation setup.

\subsection{Modeling open boundaries and ramps}
\label{sec:boundariesAndRamps}
The modeling of boundaries and ramps still offers several degrees of freedom and can significantly influence 
the resulting traffic patterns. Apart from the already mentioned boundary-induced phase transitions~\cite{Krug1991}, 
the boundary setup can affect and limit the spectrum of observable traffic states~\cite{BarlovicHuisingaSchadschneiderSchreckenberg2002}.

Therefore, and for better reproducibility of our results, we want to give a concise description of the boundary and on-ramp setup. 
To insert new vehicles, we define an entrance section of length $l=x_\mathrm{down}-x_\mathrm{up}$, 
where $x_\mathrm{down}~(x_\mathrm{up})$  stands for the
downstream (upstream) end of the entrance section. 
A vehicle entering the system adapts its velocity to the average velocity of the following and leading vehicles. The new vehicles is placed in the largest gap of the
entrance section if the insertion is safe. The insertion is safe if neither the inserted vehicle's deceleration nor the 
following vehicle's deceleration falls below a threshold $b_\mathrm{th} = \SI{-1}{\meter\per\square\second}$. 
(Due to the longitudinal-transversale coupling one has to check the other lane as well in the case of the IDM.)
This strategy can be applied both to the on-ramp and to the upstream boundary.

Similarly, vehicles leaving the road via the off-ramp are selected from an exit section. The insert and exit sections 
representing the on-ramp and the off-ramp are restricted to the rightmost lane, whereas the upstream boundary's entrance section 
spans both lanes.

To mimic the traffic state at the downstream boundary, we apply a dynamic speed limit from the position of detector 
D10 to the end of the road. The speed limit equals the maximum velocity that detector D10 measured during the 
previous aggregation interval. (As the CA models require an integer value for the speed limit, we converted the 
value by rounding it up.) 
To allow for a smooth deceleration with respect to the speed limit, we defined 
a deceleration section upstream of detector D10. 
Vehicles within the deceleration section gradually reduce their velocity such that they
pass detector D10 with the desired velocity.

Vehicles are inserted according to the empirical data of detector D01. Consequently, the open boundary's entrance 
section starts at the position of D01. For the length of the entrance section we used a value of $\SI{112.5}{\meter}$, 
which is half the value we used for the on-ramp's (off-ramp's) entrance (exit) section. 
As there are no detectors directly measuring the vehicles entering or leaving the road via ramps, we determine these values 
by comparing the values of detectors D07 and D09 with each other. D07 (D09) is situated immediately downstream 
(upstream) of the off-ramp (on-ramp). 
If detector D09 measured more vehicles than D07, the surplus is inserted in the on-ramp section. 
Otherwise, the additional vehicles are removed from the exit section. 
In principle, it would also be possible to treat the D07- and D09-detector data separately. 
In this approach, however, we might remove a vehicle from the exit section and insert a new one in the entrance
section at the same time. 
As each insertion represents a perturbation of traffic flow, which we want to minimize, 
this approach is less favorable.
If an insertion or deletion fails, it is retried until it succeeds in conserving the total number of vehicles. 

Moreover, trucks were not allowed to overtake in our simulation for two reasons. 
First, with asymmetric lane changing rules slower vehicles are required to stay in the right lane by legislation. 
Second, the examined road section is equipped with dynamic traffic signs which may impose an overtaking 
ban for trucks if traffic demand is high. 
Consequently, trucks were inserted only in the right lane at the upstream boundary. 
Figure~\ref{exemplary-time-series} supports this assumption because the number of trucks detected in the left lane 
is quite low compared with the right lane.

\subsection{Calibration}
In contrast to previous works~\cite{BrockfeldKuehneSkabardonisWagner2003,BrockfeldWagner2005,PunzoSimonelli2005}, 
we tried to avoid an automated model calibration, as an automated calibration not only 
questions the reliability of the obtained values~\cite{BrockfeldKuehneSkabardonisWagner2003} but can also produce 
parameter sets whose degree of realism can sometimes be doubted (see, \eg the optimized parameter sets in \cite{PunzoSimonelli2005}).

\begin{figure}
	\centering
	\includegraphics[width=.8\textwidth]{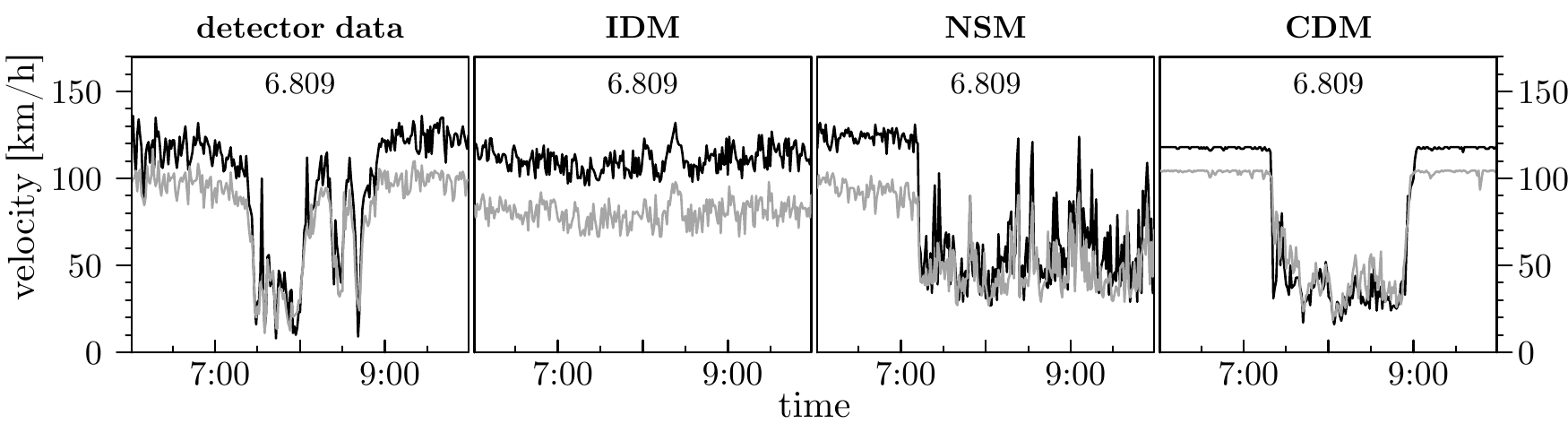}
	\caption{The empirical time series (left) and the time series produced by the three models with the original sets of parameters.}
  \label{fig:unmodified}
\end{figure}

Due to various factors (\eg weather, road characteristics, or different drivers), one cannot forgo a calibration 
completely though. As it might be quite instructive, we will briefly discuss the calibration steps we took. For the necessity 
of a calibration see figure~\ref{fig:unmodified}, which shows the actual time series at detector D07 and the 
velocity time series of the uncalibrated models at the same position. As one can see, the NSM overestimates the severity of the breakdown, 
whereas the IDM does not show a breakdown at all. 
In the case of the NSM the calibration is obvious as the model has only one parameter left to calibrate 
(see table~\ref{table-model-compare}). This parameter controls the strength of fluctuations in traffic flow. Hence,
we slightly reduced its initial value from $0.4$ to $0.32$. Despite its larger number of parameters, a calibration of
the IDM was straightforward as well: we increased a driver's desired time headway from \SI{1.2}{\second} to \SI{1.5}{\second}. 
Thereby, we reduced the road's effective capacity. 
(This change is still in good agreement with empirical data~\cite{KestingTreiber2008}.) In this case, however, we found 
that many drivers were no longer able to enter the main road via the on-ramp due to the previously defined safety criteria. 
Therefore, we changed the threshold of the acceptable deceleration of following vehicles to $b_\mathrm{th} = \SI{-50}{\meter\per\square\second}$. 
(Higher values of $b_\mathrm{th}$ did not yield satisfying results.)

\section{Results}
\label{sec:results}
\begin{figure}
	\centering
	\includegraphics[width=.85\textwidth]{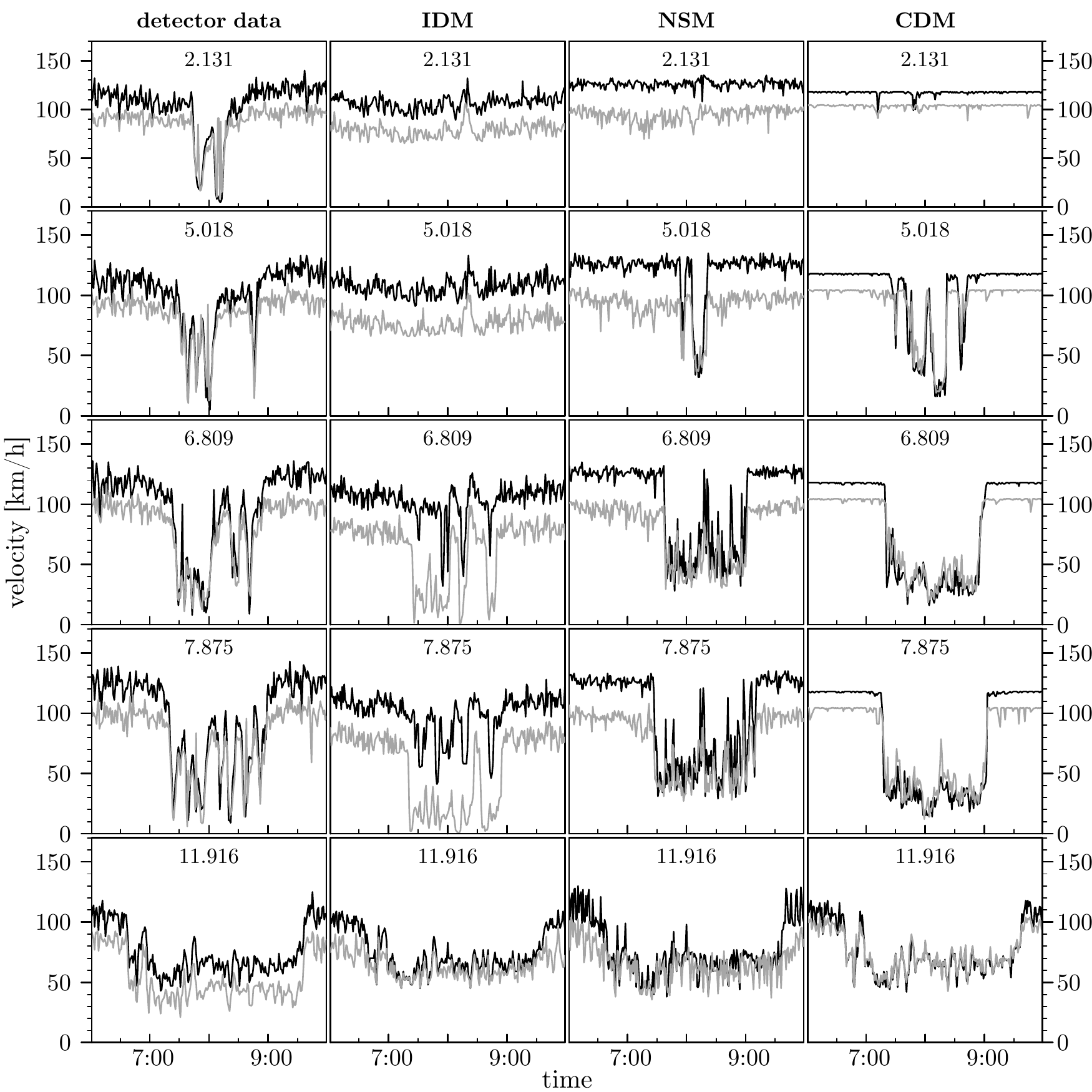}
	\caption{A comparison between empirical time series (first column) and the corresponding ones obtained with 
	the IDM, NSM, and CDM (second to fourth column), respectively. From top to bottom the rows show the time 
	series for D02, D04, D05, D06, and D10. The average velocities for the left (right) lane are plotted with a 
	black (gray) line. Please note that, for the sake of clarity, we limited the depicted time to the interval 
	from 6~a.m. to 10~a.m. Our analysis, however, is based on the time series of the entire day.}
  \label{fig:time-series-matrix}
\end{figure}

A first evaluation of the models' behavior is best carried out by a graphical inspection of the resulting time series. 
Apart from the empirical time series, figure~\ref{fig:time-series-matrix} shows the corresponding time series from several 
detector cross-sections for all models during the morning peak hour. 
The empirical time series of detectors upstream of the bottleneck (\ie D02 -- D07) show a breakdown between 7:15~a.m. 
and 9:00~a.m. During this period, the average velocity on the faster left lane abruptly drops below the free flow velocity on 
the slower right lane and the velocity synchronizes across lanes. The temporal extent as well as the severity of the 
breakdown decreases with growing distance to the bottleneck (D06 $\rightarrow$ D02). Due to boundary effects 
detector D10 also shows a velocity synchronization with weaker fluctuations and a higher average velocity.

All models' time series show good agreement with the empirical data at the downstream boundary (detector D10). 
This is, of course, a consequence of the dynamic speed limit which is imposed on the downstream (see section~\ref{sec:boundariesAndRamps}). 
Upstream of the bottleneck region, however, deviations between the models become 
obvious. The IDM's time series does not show a synchronization between the two lanes. Although one can observe a breakdown 
in either lane, traveling in the left lane is always faster than in the right lane. 
This is a result of the very aggressive attitude ($b_\mathrm{th} = \SI{-50}{\meter\per\square\second}$) 
with which drivers from the on-ramp enter the right lane of the main road. Greater values of $b_\mathrm{th}$, however, 
resulted in extinction of the breakdown. In contrast, both cellular automaton models reproduce the lane 
synchronization during traffic breakdown even for $b_\mathrm{th} = \SI{0}{\meter\per\square\second}$. 
This follows from the models' weaker safety constraints, which also result in the very unrealistic 
deceleration behavior observed in figure~\ref{fig:velocity-profile}.

Another fundamental difference is the origin of velocity fluctuations in free traffic flow: 
the CDM and NSM incorporate random fluctuations by a stochastic component, whereas the fluctuations in the IDM result 
from the heterogeneity of the traffic (see table~\ref{table-model-compare}). 
For the first two models the strength of velocity fluctuations is governed by their spatial discretization, 
which explains why the fluctuations in the CDM's time series are less pronounced than for the NSM. 

As one can see from figure~\ref{fig:time-series-matrix}, the duration of the breakdown decreases in the upstream direction  
in both the empirical data and the data generated by the traffic models. All models underestimate the breakdown's spatial extent though.
At detector D02, free flow is restored in all models whereas empirical data still show the 
influence of the breakdown.

\begin{figure}[htb]
\centering
\subfigure[\label{fig:acfIDM} IDM]{\includegraphics[width=0.32\textwidth]{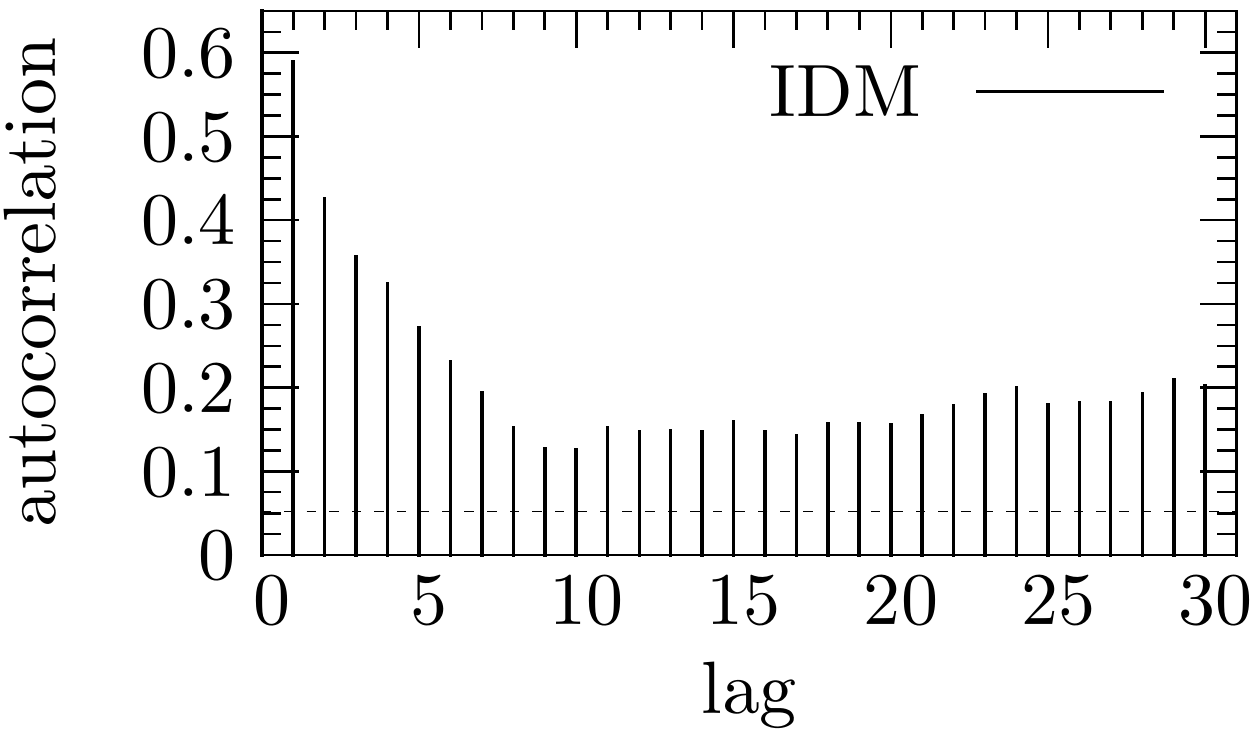}}
\subfigure[\label{fig:acfNASCH}NSM]{\includegraphics[width=0.32\textwidth]{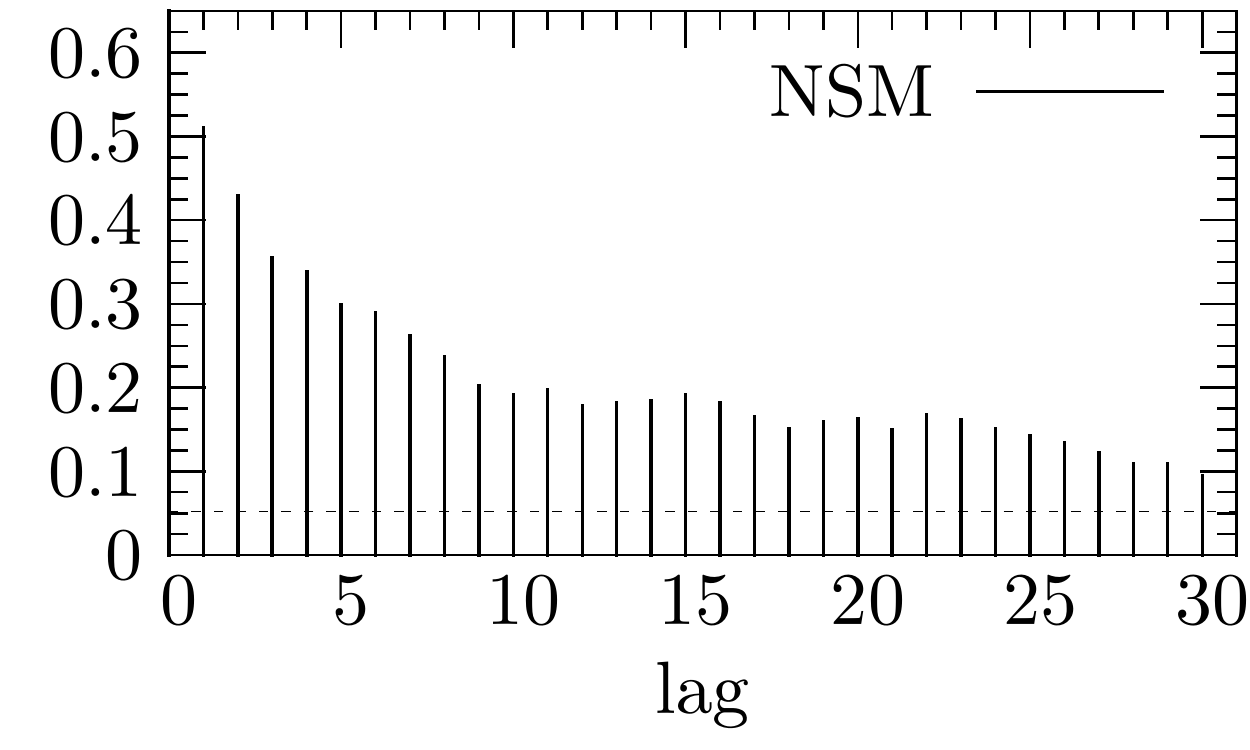}}
\subfigure[\label{fig:acfCDM} CDM]{\includegraphics[width=0.32\textwidth]{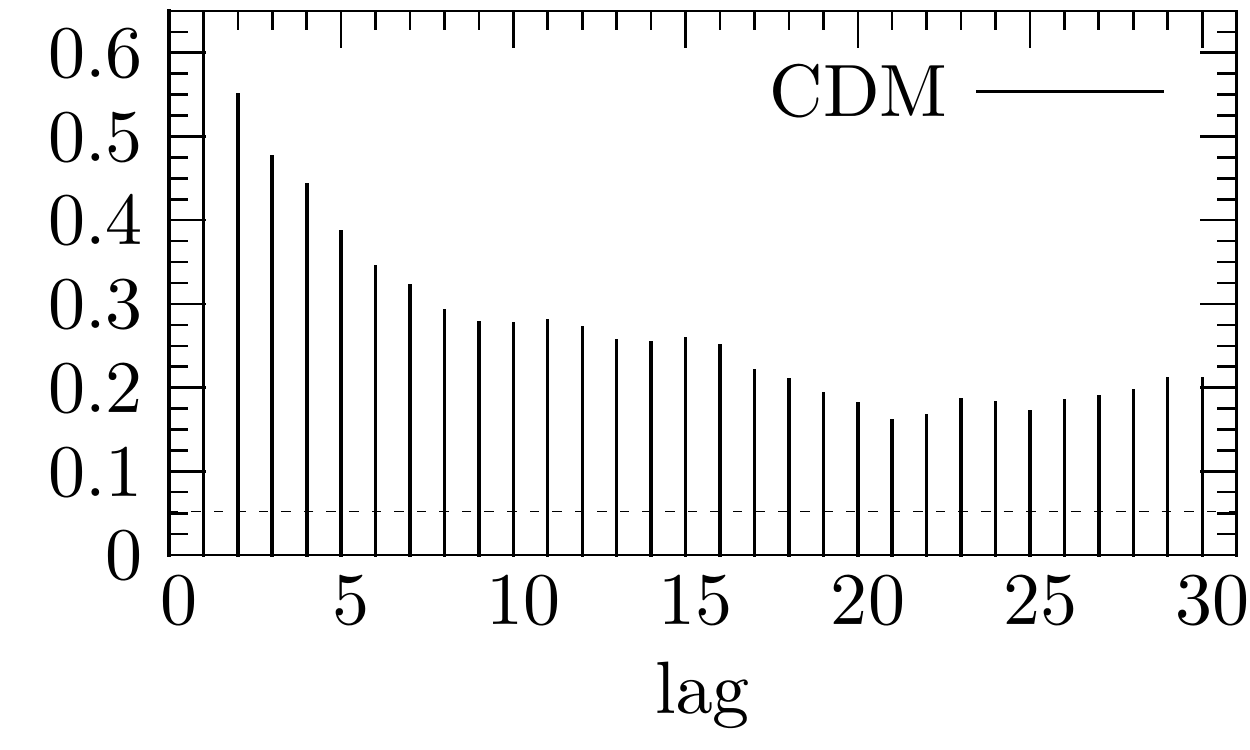}}\\
\subfigure[\label{fig:correlogram} lane averaged correlation coefficients]{\includegraphics[width=0.42\textwidth]{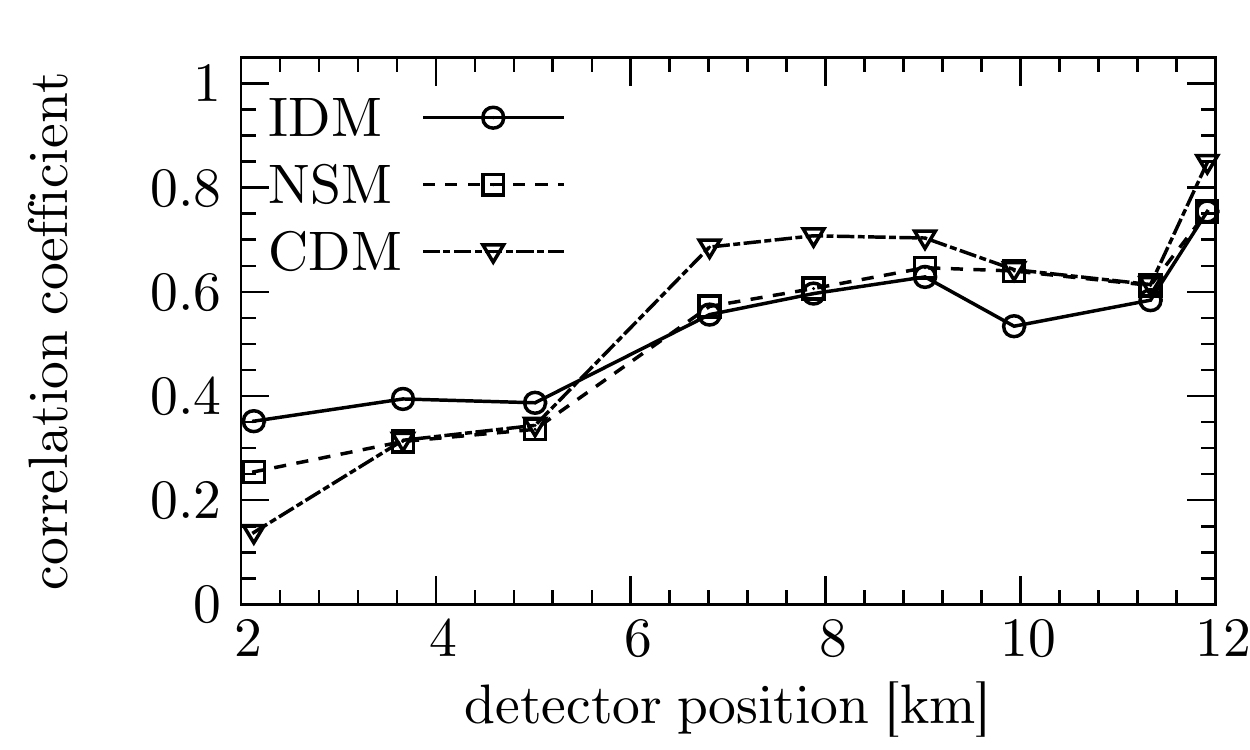}}
\subfigure[\label{fig:affine-error}1-norm]{\includegraphics[width=0.42\textwidth]{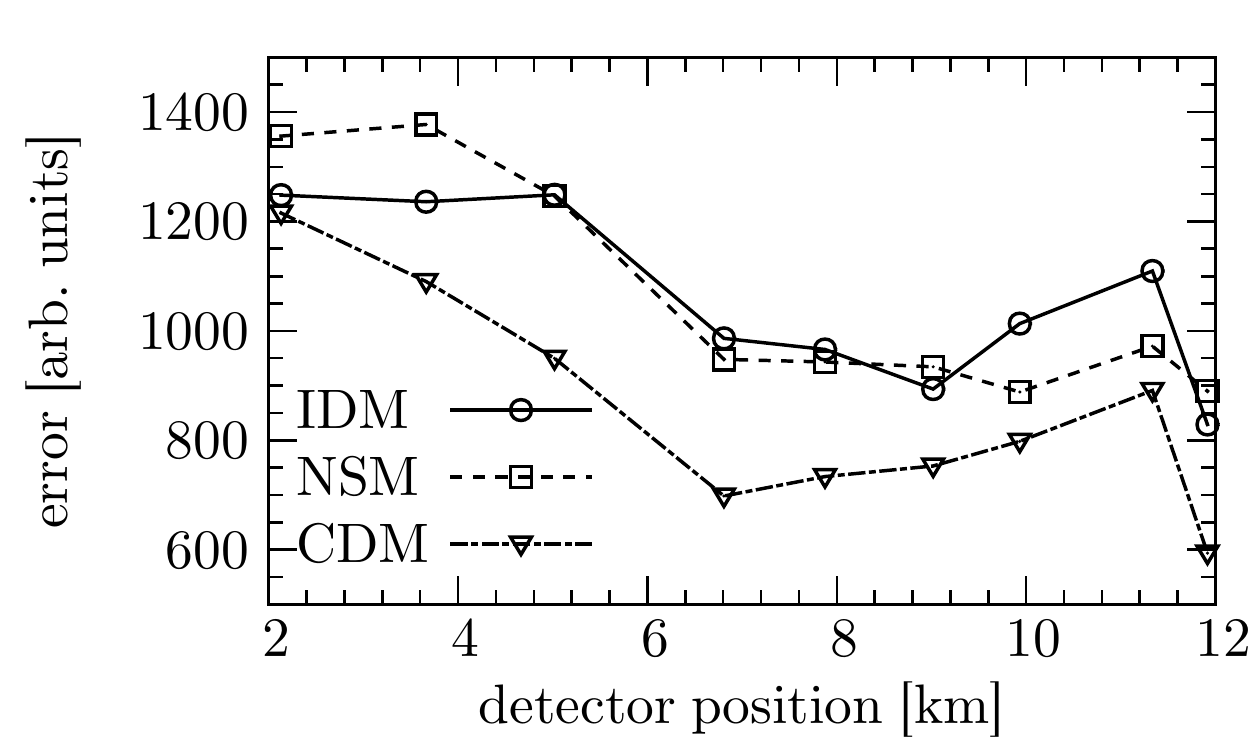}}
\caption{\label{fig:compare}Autocorrelation of the time series' residuals at detector D05 for the IDM 
		\subref{fig:acfIDM}, NSM \subref{fig:acfNASCH}, and CDM \subref{fig:acfCDM}. Assuming white noise 
the $95\%$-confidence interval is plotted as a dashed horizontal line. 
The cross correlation between the empirical time series and the models' time series is shown in \subref{fig:correlogram}. 
The 1-norm according to equation (\ref{eqn:1norm}) is given in \subref{fig:affine-error}.}
\end{figure}

Similarly to the graphical inspection the numerical analysis does not reveal qualitative differences in the 
models' velocity time series. 
Figures~\ref{fig:acfIDM}--\subref{fig:acfCDM}  show the residuals' autocorrelation for each model 
at detector D05, which is representative for the other detectors upstream of the on-ramp. 
Ideally, the residuals are uncorrelated, yielding an autocorrelation coefficient close to zero. 

All models show a positive correlation ($>0.5$) for the residuals, which decreases with increasing time-lag. 
Hence, the assumption of white noise for the residuals has to be rejected.
Figure~\ref{fig:correlogram} shows the correlation between the models' time series and the empirical data. 
Due to the imposed downstream boundary conditions, the correlation coefficient's maximum value ($> 0.7$) 
is obtained  at D10  for all models.
For all models the correlation coefficient remains above \num{0.5} up to detector D05 indicating a satisfying 
reproduction of the empirical data. 
The decrease in correlation with increasing distance to the ramps reflects the fact that all models do not successfully 
reproduce the spatial extent of the breakdown, as already discussed.


Probably the most intuitive comparison is obtained by the calculation of the 1-norm after transforming the time series 
according to equation (\ref{eqn:affine-transformation}). Remember that our  primary 
quality criterion is the reproduction of the breakdown. 
Hence, we expect it to indicate to what extent the corresponding models achieve this goal. 
Figure~\ref{fig:affine-error} shows the 1-norm calculated for each detector and averaged over both lanes. 
At D02 (\ie at kilometer $2.131$) the error  for all models is about $1300$ (measured in arbitrary units) 
as all models fail to reproduce the breakdown at this position. 
Again, due to the boundary conditions, the best accordance with the empirical measurements is observed at detector D10. 
The qualitative behavior between detectors D01 and D10 is similar for all detectors. 
Even the absolute values of the NSM and the IDM are nearly 
identical; only the CDM's values are considerably below the values of the previous models and indicate a better 
performance of this model.

\subsection{Reproducibility of upstream boundary conditions}
As we have seen, all investigated models are, in principle, able to reproduce the spatiotemporal traffic dynamics. 
As the simulations used the real detector data, it is worth discussing to what extent the empirical boundary conditions 
could actually be reproduced. 
In section~\ref{sec:boundariesAndRamps} we described how vehicles were inserted and that a failed insertion had to be repeated in subsequent 
time steps until it succeeded.
Of course, one would expect an insertion to always be successful as the empirical data prove that 
real traffic can satisfy the observed demand. 
In the simulations, we saw that insertions repeatedly failed especially during the peak hour period: 
table~\ref{table:queues} gives the number of cars waiting for insertion or removal at the on-ramp or the off-ramp, respectively. 
Moreover, we have recorded the maximum time for either ramp during which the queue of vehicles waiting for insertion or removal was 
above zero. 
For instance, the maximum number of vehicles waiting for insertion in the NSM was 63 and the time span of a non-empty queue was 
\SI{14}{\minute}.
\begin{table}[htb]
\caption{Overview of the usage of the on- and off-ramp queues}\label{table:queues}
\centering
\begin{tabular}{lrrrr}
\toprule
      & \multicolumn{2}{c}{max. queue length [veh] at} & \multicolumn{2}{c}{max. queue duration [\si{\minute}] at}\\
      \cmidrule(r){2-3}\cmidrule(r){4-5}
model & on-ramp & off-ramp & on-ramp& off-ramp\\
\midrule
IDM   & 139 & 41 & 82 & 22\\
NSM   & 63  & 42 & 14 &  9\\
CDM   & 55  & 42 & 10 & 41\\
\bottomrule
\end{tabular}
\end{table}
These results confirm our previous statement that the safety conditions of the IDM make vehicle insertion more difficult. 
The maximum queue length for the IDM is twice as large as those for the two cellular automaton models.
For all models, the queue length at the on-ramp reaches values considerably above zero.
Whether a vehicle insertion fails depends on each model's safety conditions. 
Moreover, the queue length's growth reflects, at least partially, the fact that the models do not perfectly mimic human driving behavior.
It has been found found, for example, in real traffic that the time headway of vehicles traveling at 
\SI{110}{\kilo\meter\per\hour} can even drop below \SI{0.5}{\second} \cite{AppertRoland2009}.
Such values are impossible for all models due to their safety constraints.
(In real traffic, such headways pose an actual risk, as drivers cannot react to actions of the preceding vehicle in a timely manner.)

\subsection{Propagation velocity of breakdown}
Another characteristic property of traffic flow is the propagation velocity 
of traffic jams and the shock (see section \ref{sec:related_work}). 
Figure~\ref{fig:time-series-matrix} shows that the breakdown propagates upstream.
As the upstream and downstream fronts of a traffic jam are associated with sharp discontinuous changes 
of the average velocity, we define the time when the shock reaches the detector as the 
time when the average velocity first drops below a given threshold $v_\mathrm{th}$. 
Similarly, the time when the average velocity first exceeds $v_\mathrm{th}$ is considered as the detection of  
the downstream jam front. 
The numerical values of the propagation velocity of the shock and the traffic jam can be obtained 
from the quotient of the distance between subsequent detectors and the time that passed between 
the events being registered at either detector.
In accordance with \cite{RehbornKlenovPalmer2010}, we have set $v_\mathrm{th}=\SI{30}{\kilo\meter\per\hour}$.
As the temporal resolution of one minute is relatively coarse, we averaged over the interval 
$v_\mathrm{th} =$~\SI[separate-uncertainty = true]{30+-3}{\kilo\meter\per\hour} (see table \ref{table:breakdownVelocity}).
\begin{table}[htb]
	\caption{Averaged breakdown propagation velocities. 
	The velocities are given in units of \si{\kilo\meter\per\hour}.}
	\label{table:breakdownVelocity}
	\centering
	\begin{tabular}{lrrrrrrrr}
	\toprule
    	  & \multicolumn{4}{c}{shock-wave velocity for} & \multicolumn{4}{c}{jam propagation velocity for}\\
	      \cmidrule(r){2-5}\cmidrule(r){6-9}
	detectors & real data & NSM & IDM & CDM & real data & NSM & IDM & CDM\\
	\midrule
	D07$\rightarrow$D06 & -23 & -7  & -5  & -2 & -15 & -17 & -24 & -9\\
	D06$\rightarrow$D05 & -12 & -32 & -14 & -7 & -11 & -28 & -3 & -7\\
	D05$\rightarrow$D04 & -11 & -3  & NA  & -3 & -13 & -3 & NA & -3\\
	\bottomrule
\end{tabular}
\end{table}
Although the propagation velocities vary strongly between models and between successive detectors, 
all models correctly predict a negative propagation velocity (\ie the breakdown moves upstream
against the direction of traffic flow). 
Due to the relatively small inter-detector distances and due to an approximate error of \SI{30}{\second} in the 
temporal determination of the breakdown, the error margins are quite large though. 
In principle, the error can be decreased by considering two detectors far away from each other 
(\eg D07$\rightarrow$D02). 
For the empirical data set we found the propagation velocity to be \SI{-13}{\kilo\meter\per\hour} 
for D07$\rightarrow$D02 and D06$\rightarrow$D02, which is in good agreement 
with the value of \SI[separate-uncertainty = true]{-15+-2}{\kilo\meter\per\hour} 
found by Rehborn and colleagues~\cite{RehbornKlenovPalmer2010}.

\subsection{Influence of truck length and the downstream boundary condition}
\begin{figure}
	\centering
	\includegraphics[width=.85\textwidth]{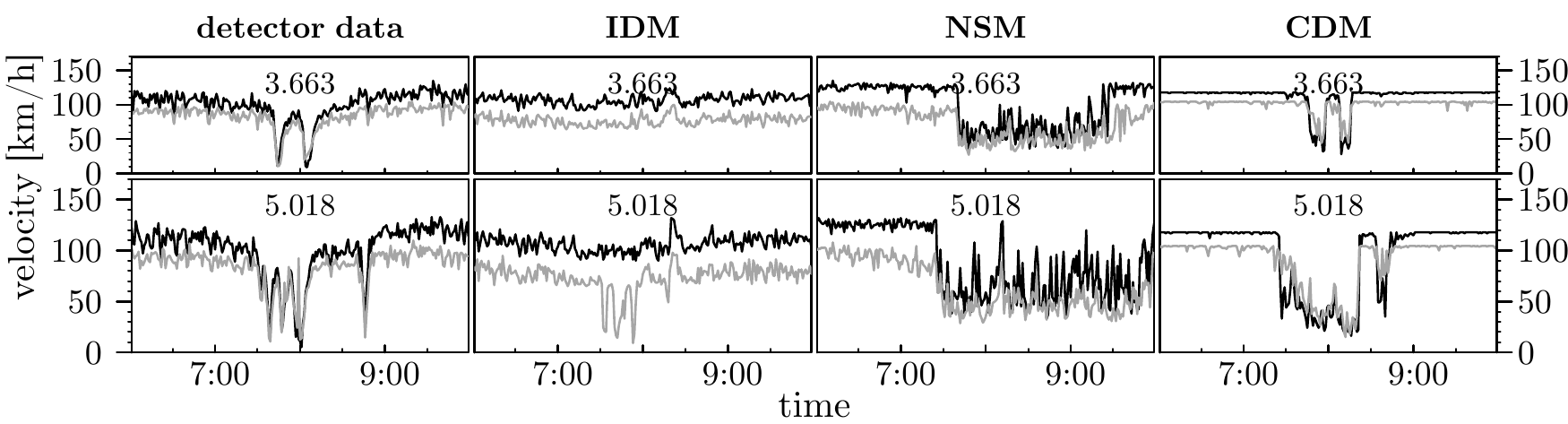}
	\caption{After increasing the length of trucks, the breakdown can be observed at detectors further upstream 
	 (cf. figure~\ref{fig:time-series-matrix}).}
    \label{fig:trucklength}
\end{figure}

\begin{figure}
	\centering
	\includegraphics[width=.85\textwidth]{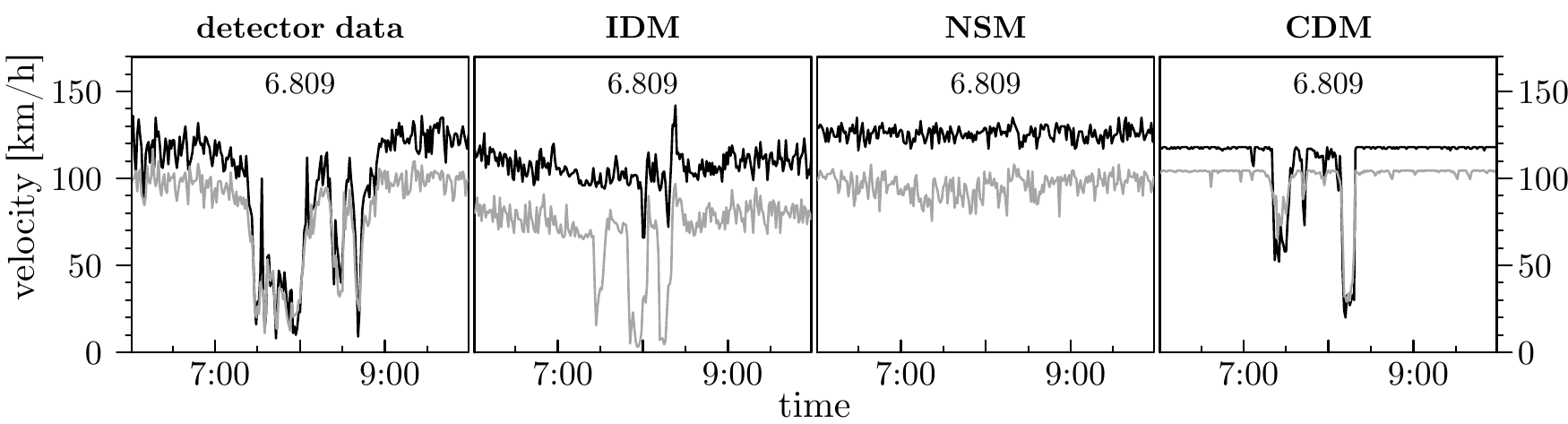}
	\caption{When modeling free flow at the downstream boundary, the intensity of the breakdown decreases
	in all models. For the NSM the breakdown even completely disappears.}
    \label{fig:withoutspeedlimit}
\end{figure}

Finally, we want to demonstrate the influence of vehicle length and the downstream boundary on our results. 
The vehicle length directly influences the spatial extent of the breakdown. 
Figure~\ref{fig:trucklength} shows the velocity time series from simulation runs where we increased the length of trucks by 
\SIlist{7.5;1;1.5}{\meter} for the NSM, the IDM, and the CDM, respectively.
A comparison with figure~\ref{fig:trucklength} proves that this slight change of the vehicle properties 
results in a better agreement with the empirical data.
For the NSM not only the spatial extent but also the temporal duration of the breakdown increases. 
Due to the NSM's spatial discretization vehicle length can only be altered by multiples of \SI{7.5}{\meter}.
A more fundamental aspect is the nature of the breakdown. On the downstream boundary we had imposed a variable speed limit 
to mimic the velocities observed in real traffic. 
Hence, it is worth investigating whether the breakdown resulted from the imposed speed limit or whether it resulted from 
the vehicle interactions in the on-ramp region. 
Therefore, we repeated the simulations without the speed limit and, thereby, guaranteed  
free flow at the downstream boundary. 
The resulting exemplary time series for detector D06 is depicted in figure~\ref{fig:withoutspeedlimit}. 
The NSM's time series no longer exhibits a breakdown, whereas the IDM and CDM still do. 
Consequently, the source of the breakdown in the NSM was our choice of the downstream boundary's setup, 
whereas the breakdowns in the CDM and IDM are caused by the on-ramp boundary. 
However, also for the latter two models the shape of the time series changed: the breakdown's duration decreased. 
Therefore, even if it does not select the traffic state (congested or freely flowing), the downstream boundary does determine its duration.

\section{Conclusion}
\label{sec:conclusion}
In this paper, we investigated the ability of three microscopic traffic models, namely 
the Nagel-Schreckenberg model (NSM), the comfortable driving model (CDM), 
and the intelligent driver model (IDM), to reproduce 
spatiotemporal traffic dynamics. To do so, we modeled a section of a 
German Autobahn with an off-ramp and an on-ramp. 
The inflow rates (via the upstream boundary and the on-ramp) and outflow rates (via the off-ramp) 
of vehicles were determined by the empirical traffic data that show a spontaneous breakdown during the morning peak hours.
To assess our results and to compare the models with each other, we employed 
several methods known from time series analysis. 
We found that all models can reproduce the breakdown satisfyingly---at least 
after minor calibrations. 
Although the three models investigated here are microscopic ones, our observation is 
quite surprising as there are still substantial differences between the models 
(\eg continuous versus discrete in space, realistic versus unlimited braking capacities, 
or deterministic versus stochastic approach).

Similarly, all models showed the same shortcomings. Compared with the empirical data, they underestimated 
the spatial extent of the breakdown. 
In real data, the breakdown could be detected more than \SI{8}{\kilo\meter} upstream of the on-ramp, which 
none of the models could reproduce. 
Therefore, we investigated the influence of the vehicles' length on the traffic dynamics.
We saw that the spatial extent of the breakdown can be effectively adjusted by the lengths of the different vehicle types. 

Adjusting the length of vehicles, whose number is given by the boundary conditions, has a direct impact on 
the vehicle density and, thereby, on the probability with which vehicles can enter the road from the on-ramp.
In this context, we did observe qualitative differences between the models though: 
in peak-hour traffic (\ie high vehicle density), vehicles could more easily enter the main road via the on-ramp in the NSM and the CDM
than in the IDM. 
The better reproduction of the empirical inflow rates exhibited by the CDM and NSM is a direct consequence of 
their unlimited braking capacity. The IDM, which shows a realistic deceleration behavior, makes higher demands on a 
safe entry and, therefore, it is more difficult for vehicles to enter in high density phases. 

In contrast to the CDM and IDM, we found that the breakdown's occurrence in the NSM did strongly depend on the modeling 
of the downstream boundary. Without applying a speed limit to the downstream boundary, we could no longer observe a breakdown in the NSM. 
From a physical point of view this is major difference between models.  
The consequences for practical applications, however, are less severe:
for large scale traffic simulations (\eg \cite{HafsteinChrobokPottmeierSchreckenbergMazur2004, Olsim}),
the downstream boundary conditions are given by the traffic state of the neighboring road segment and cannot 
be chosen arbitrarily. Therefore, if the primary concern is the reproduction of the current traffic state, 
its physical origin may be of secondary importance.

As all models gave good results compared with the empirical data, is not possible to make a definite recommendation 
for one of them. 
The highest simulation speeds were observed with the NSM. 
Security-related aspects, which require a realistic acceleration and deceleration behavior, are best answered with the IDM. 
If the focus is rather on a realistic description of the spatiotemporal traffic dynamics, then the CDM is a good alternative. 
The authors, for example, recently studied the influence of inter-vehicle communication on traffic flow with the help of the CDM 
\cite{KnorrSchreckenberg2012,KnorrBaseltSchreckenbergMauve2012}. 
In this context, the CDM offered a good combination for fast and realistic traffic simulations.
\section*{Acknowledgments}
FK would like to thank A~Kesting and M~Treiber for discussing various aspects of 
the IDM with him.
FK's work was funded by the state of North Rhine--Westphalia and the European Regional 
Development Fund (ERDF) within the NRW--EU Ziel 2 program ``automotive.nrw''.

\section*{References}
\bibliographystyle{unsrt}
\bibliography{literaturverzeichnis}

\end{document}